\documentclass[apj]{emulateapj}
\usepackage{apjfonts}
\usepackage{color}
\newcommand\beq{\begin{equation}}
\newcommand\eeq{\end{equation}}

\newcommand{\chandra}{\textit{Chandra}}

\shorttitle{Baryon Content in Massive Clusters}
\shortauthors{Lin et al.}

\slugcomment{ApJL, in press}

\begin{document}

\title{Baryon Content of Massive Galaxy Clusters at \lowercase{$z=0-0.6$}}

\author{
Yen-Ting Lin\altaffilmark{1},
S.~Adam Stanford\altaffilmark{2},
Peter R.~M.~Eisenhardt\altaffilmark{3},
Alexey Vikhlinin\altaffilmark{4},
Ben J.~Maughan\altaffilmark{5},
and Andrey Kravtsov\altaffilmark{6}
}

\altaffiltext{1}{Institute of Astronomy and Astrophysics, Academia Sinica, Taipei, Taiwan; Institute for the Physics and Mathematics of the Universe, Todai Institutes for Advanced Study, The University of Tokyo, Kashiwa, Chiba, Japan; ytl@asiaa.sinica.edu.tw}
\altaffiltext{2}{Physics Department, University of California, Davis, CA 95616; Institute of Geophysics and Planetary Physics, Lawrence Livermore National Laboratory, Livermore, CA 94551}
\altaffiltext{3}{Jet Propulsion Laboratory, California Institute of Technology, Pasadena, CA 91109}
\altaffiltext{4}{Harvard-Smithsonian Center for Astrophysics, Cambridge, MA 02138} 
\altaffiltext{5}{HH Wills Physics Laboratory, University of Bristol, Bristol, UK} 
\altaffiltext{6}{Department of Astronomy and Astrophysics, Kavli Institute for Cosmological Physics, and Enrico Fermi Institute, The University of Chicago, Chicago, IL 60637}

\begin{abstract}

We study the relationship between two major baryonic components in galaxy clusters, namely the stars in galaxies, and the ionized gas in the intracluster medium (ICM), using 94 clusters that span the redshift range $0-0.6$. Accurately measured total and ICM masses from \chandra\ observations, and stellar masses derived from the Wide-field Infrared Survey Explorer and the Two-Micron All-Sky Survey  
allow us to trace the evolution of cluster baryon content in a self-consistent fashion.
We find that, within $r_{500}$, the evolution of the ICM mass--total mass relation is consistent with the expectation of self-similar model, while there is no evidence for redshift evolution in the stellar mass--total mass relation.  This suggests that the stellar mass and ICM mass in the inner parts of clusters evolve differently.

\end{abstract}

\keywords{galaxies: clusters: general -- galaxies: elliptical and lenticular, cD -- galaxies: luminosity function, mass function -- galaxies: clusters: intracluster medium}

\section{Introduction}
\label{sec:intro}

The baryon content in groups and clusters of galaxies is one of the basic quantities that could be used to test models of structure formation.
The total baryon mass fraction in massive clusters has been used to estimate cosmological parameters \citep[e.g.,][]{swhite93a,allen04}.  
The mass of stars in galaxies is a fossil record of the star formation and galaxy merger history \citep{conroy07}.
The relative proportion in mass of hot (diffuse intracluster medium, ICM) and cold baryons (e.g., stars in galaxies) 
could be used to gauge the strengths of various feedback mechanisms and star formation efficiency \citep[e.g.,][]{bode09}.

Although many measurements of the baryon fraction in local clusters exist \citep[e.g.,][]{lin03b,gonzalez07,andreon10,balogh11}, 
such studies using large cluster samples at higher redshifts are in short supply.
The first attempt has been made with the X-ray selected groups at $z=0.1-1$ detected in the COSMOS field \citep[][hereafter G09]{giodini09}.  However, the total and ICM masses are not well-measured for individual systems, so G09 could not constrain $M_{\rm ICM}/M_{\rm star}$ well, especially at the massive end (see also \citealt{leauthaud11}).

Here we present a measurement of the baryon mass fraction contained in galactic stars and ICM, as well as the relationship between $M_{\rm ICM}$ and $M_{\rm star}$, using 
94 clusters at $z\le 0.6$ -- 
the largest sample with accurate total and ICM mass measurements to date for this purpose.
For the $z>0.1$ systems, we use data from the Wide-field Infrared Survey Explorer (WISE; \citealt{wright10}) survey to determine $M_{\rm star}$, while for the local clusters, we rely on data from the Two-Micron All-Sky Survey (2MASS, \citealt{skrutskie06}).
To ensure uniformity of our measurements and to facilitate comparison across the cosmic epochs, all quantities are measured within $r_{500}$, the radius within which the mean overdensity is 500 times the critical density of the Universe $\rho_c$ at the cluster redshift.

In this paper we have neglected neutral and molecular gas in the baryon budget, as their amount is believed to be small \citep[e.g.,][]{chung09}.
Our analysis does not include the intracluster stars either (as they are well below the surface brightness limit of WISE), although they may contribute a non-negligible fraction to the total cluster stellar mass \citep[e.g.,][]{gonzalez07}.
Throughout this paper we adopt a {\it WMAP-5} \citep{komatsu09} $\Lambda$CDM 
cosmological model where $\Omega_M=1-\Omega_\Lambda=0.26$,
$H_0=71 h_{71}\,{\rm km\,s}^{-1} {\rm Mpc}^{-1}$.

\section{Analysis Overview}
\label{sec:overview}

Our analysis relies on the X-ray measurements to provide the cluster center, size, and the mass of the ICM.  The cluster samples used are described in \S\ref{sec:samples}.  
We discuss the way artifacts and stars are rejected from the WISE catalogs
in \S\ref{sec:wise}.  Throughout this study we only make use of WISE 3.4 $\mu$m data.
For each cluster, we use the background-subtracted total flux to estimate the total cluster luminosity and stellar mass.  The method is described fully in \citet{lin03b,lin04}.  In \S\ref{sec:stellarmass} we outline our modified procedures for analyzing the WISE data.

\subsection{Cluster Samples}
\label{sec:samples}

We assemble our intermediate redshift ($z=0.1-0.6$) cluster sample from \citet[][hereafter M08]{maughan08} and \citet[][hereafter V09]{vikhlinin09}, both of which are based on \chandra\ observations.
Although the 115 clusters from M08 are a heterogeneous sample (selected to be targeted ACIS-I observations, at $z>0.1$, publicly available in \chandra\ archive as of 2006), the large sample, analyzed in a uniform fashion, is advantageous, and allows us to probe the baryon content in clusters of various dynamical states.  
The \chandra\ subsample of the ROSAT PSPC 400 deg$^2$ survey presented in V09, on the other hand, has a well-defined selection function, but is smaller in size (41 objects).
We have taken from these studies the X-ray center, ICM and total mass\footnote{For M08 clusters, the measurements are revised with \chandra\ CALDB version 4.3.1.  For V09 clusters, all the measurements presented here use cir.~2005 calibration; with the most recent calibration, they change only by $\sim 3\%$.}. The total mass $M_{500}$ is derived from the $Y_X$--$M_{500}$ relation \citep{kravtsov06}, where $Y_X$ is the product of core-excised X-ray temperature $T_X$ and $M_{\rm ICM}$.
Using clusters common in the two samples we have verified that the two analyses yield measurements consistent to $\lesssim 5\%$.
Requiring that the clusters are fully covered in the WISE Preliminary Data Release (PDR) footprint (see below), {\it and} are not affected by nearby bright stars, our final sample consists of 45 clusters, with 29 and 16 clusters from M08 and V09, respectively.

As our local ($z<0.1$) benchmark, we combine the cluster sample presented in \citet[][hereafter L04]{lin04} with the low-$z$ systems from V09.  The resulting 49 clusters have accurate total and ICM mass measurements from \chandra\ or ROSAT, as well as total stellar mass estimates based on the near-IR $K_s$-band luminosity from 2MASS.

\subsection{WISE Data}
\label{sec:wise}

WISE is a satellite mission to survey the whole sky in four infrared bands (3.4, 4.6, 12, and 22 $\mu m$; hereafter W$1-4$), with a limiting magnitude\footnote{http://wise2.ipac.caltech.edu/docs/release/prelim/expsup/sec6\_4a.html} of 17 (Vega) in W1. 
The recent PDR has made public reduced images and source catalogs for 57\% of the sky.
For each cluster we query the NASA/IPAC Infrared Science Archive (IRSA) to download WISE PDR source catalog out to $5-10 r_{200}$, where $r_{200}$ is a proxy of the cluster virial radius, within which the mean overdensity is $200\rho_c$, and is extrapolated from $r_{500}$ by assuming an \citet[][hereafter NFW]{navarro97} density profile with concentration $c=5$.  The large area surrounding the clusters is used for estimating the local background galaxy count and flux.
In practice the background is estimated with an annulus with the inner and outer radii set to be four and five times $r_{200}$, respectively, although the exact choices vary from cluster to cluster, primarily depending on the distribution of bright star masks.  
Our results are not sensitive to the exact choice of the annulus radii.
We only use objects without any known artifacts in W1 and W2.

\begin{figure}
\epsscale{1.}
\plotone{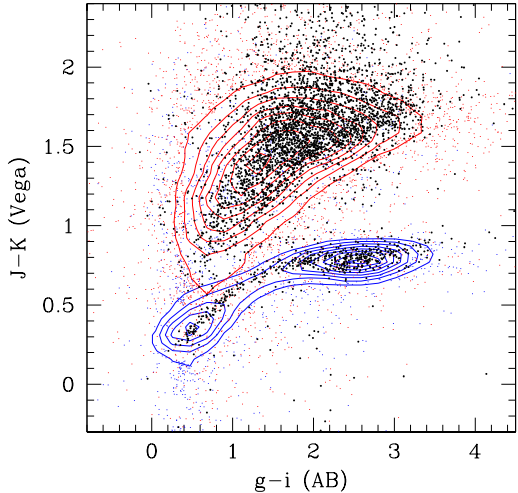}
\vspace{-5mm}
\caption{ 
The blue and red contours and points show the stellar and galaxy loci, combining data from SDSS and UKIDSS DXS.  The black points are the WISE objects without a match in the 2MASS point source catalog.  Only 7\% of such WISE sources lie along the stellar locus, which suggests that stellar contamination in our WISE galaxy catalogs is $\le7\%$.
}
\label{fig:sgsep}
\end{figure}

Because of its $6\arcsec$ FWHM point spread function (PSF), WISE does not provide useful morphological information to allow distinction between point and extended sources. 
Instead, we use 2MASS data to help select galaxies.
Specifically, for each cluster we download the 2MASS point source catalog\footnote{In addition to the standard flags used for clean photometry and to select point-like sources, we require the sources to have $J\le 15.8$.} from IRSA, covering the same area as the WISE data.  
We then regard every WISE object with a 2MASS match within $2\arcsec$ as point-like and remove them from our catalog.

To evaluate the efficiency of this approach in star-galaxy separation, we use a 1.5 deg$^2$ region located in the XMM-LSS field that has deep optical and near-IR data from CFHTLS Wide \citep{gwyn11} and UKIDSS DXS \citep{lawrence07} surveys.  In Fig.~\ref{fig:sgsep} we show the distribution of the stellar and galaxy loci on the $g-i$ {\it vs} $J-K$ diagram, obtained by cross matching the SDSS star and galaxy catalogs \citep{sdssdr8} with DXS data.
We overlay the 7163 WISE sources without a match in the 2MASS point source catalog, and find that only 7\% of them  lie on the stellar locus.
This shows that using 2MASS we can remove the great majority of the stars.  For the remaining stars in our catalogs, we remove their contribution via statistical background subtraction.

\subsection{Estimating Total Luminosity and Stellar Mass}
\label{sec:stellarmass}

\begin{figure}
\epsscale{1.}
\plotone{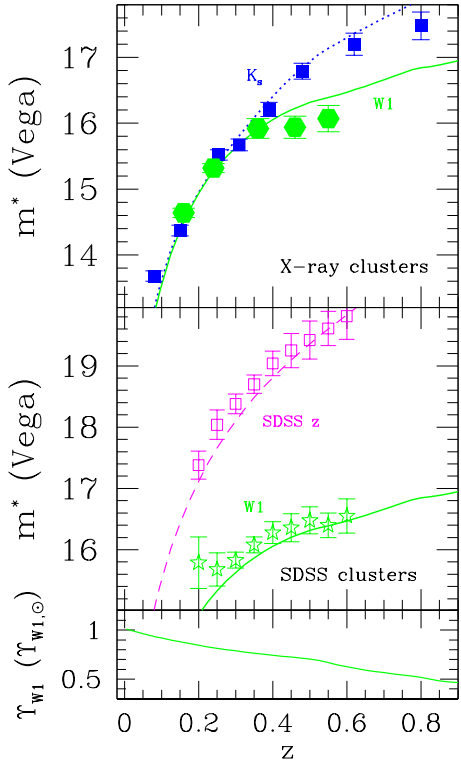}
\vspace{-4mm}
\caption{ 
Upper panel: Evolution of the apparent characteristic magnitude $m^*$ of the cluster galaxy LF in $K_s$-band (blue/solid squares; from \citealt{lin06}) and W1 (green/solid circles; our X-ray cluster sample).  
The dotted and solid curves show the predictions in $K_s$ and W1 of a single stellar population formed at $z=3$ with the Kroupa initial mass function and an exponentially decaying star formation history, normalized to fit the $K_s$-band data.
Middle panel: $m^*$ of the cluster galaxy LF in $z$-band (magenta squares) and W1 (green stars) for the optical clusters from \citet{geach11}.  
The two curves are the predictions of the same population synthesis model in $z$-band and W1. 
Lower panel: evolution of stellar mass-to-light ratio in W1.
}
\label{fig:mstarev}
\end{figure}

To describe the evolution of the cluster galaxy populations, as well as for the $k$-correction, we use a simple stellar population evolution model based on an updated version of the \citet{bruzual03} population synthesis software package (hereafter the BC model).  The model is a single stellar population of solar metallicity formed at $z=3$ with an exponentially decaying star formation history (with an $e$-folding timescale $\tau=0.1$ Gyr), with the Kroupa initial mass function (IMF), normalized 
to fit the apparent characteristic magnitude $m^*$ in the $K_s$-band for massive X-ray clusters at $z=0.1-0.8$, which are taken from \citet[][hereafter L06]{lin06}. 
Using the BC model, we follow L06 to construct the composite W1 luminosity functions (LFs) in observer's frame in five redshift bins.  For each bin, we calculate the mean redshift of the clusters $z_m$, and adjust the $k$-correction and distance modulus of individual clusters to best represent how the clusters would appear if they were all at $z_m$.
We then fit the composite LFs with the \citet{schechter76} function, and show the best-fit $m^*$ in Fig.~\ref{fig:mstarev} (top panel).

For the lower three redshift bins, the W1 $m^*$ evolution agrees well with the BC model. For the two higher redshift bins, the observed $m^*$ is about 0.3 mag brighter than the model.  To check if this is resulted from the blending of galaxies due to the large PSF of WISE, we also construct composite LFs in the SDSS $z$-band and W1 for a  sample of $\sim 900$ optically-selected clusters and groups located in SDSS stripe 82, using the catalog of \citet[][hereafter G11]{geach11}.
The $m^*$ measured for these clusters is shown in the middle panel of Fig.~\ref{fig:mstarev}, along with the BC model predictions.  
Although the measured $m^*$ of the SDSS clusters are slightly fainter than the predictions, the general agreement of $m^*$ in both W1 and $z$-band with the models indicates that blending is not the cause of the $m^*$ being brighter than the models for our X-ray cluster sample at $z > 0.4$.
The galaxies in our X-ray clusters are indeed more luminous on average than those in the SDSS optical clusters;
for example, the mean difference of the magnitudes of
the brightest cluster galaxies between the two samples is about 1 mag.
This is not surprising given that our clusters are more massive than the G11 sample on average (see e.g., L04, \citealt{hansen09}).
With the chosen BC model, our evolving $m^*(z)+2$ flux threshold (see below) corresponds to a constant galaxy stellar mass limit of $2\times 10^{10} M_\odot$.
For this reason, and
for sake of consistency, we will use the BC model for the $k$-correction and $m^*$ evolution, but will comment below on the impact of the ``brightening'' of galaxies in $z>0.4$ clusters on our results.

For each cluster, we sum the fluxes from objects brighter than the smaller of $m^*(z)+1.5$ and W1$=17.0$, and subtract fluxes from similar objects located within the background annulus (scaled by the relative area of the annulus and cluster region).
Assuming all clusters have the same shape of LF (with a faint-end power-law slope $\alpha=-0.9$), which is consistent with our data (see also L06),
we convert the background subtracted flux to $L_{{\rm tot}}$ using an NFW profile with $c=5$, defined as the total luminosity from all galaxies located within $r_{500}$ that are more luminous than $M^*(z)+2$.
On average the conversion from observed to total luminosity requires only 15\% of extrapolation\footnote{Galaxies fainter than $M^*(z)+2$ would contribute an additional 12\% in luminosity assuming $\alpha=-0.9$.}.
For reference, at $z=0$ we have $\Upsilon_{\rm 0,W1}=1.01 M_\odot/L_{\odot,{\rm W1}}$.

For our $z<0.1$ clusters, we measure the total luminosity in the $K_s$-band using 2MASS data (following L04); Again $M_{{\rm star}}=L_{{\rm tot}} \Upsilon_{K_s}(z)$, using the $K_s$-band $\Upsilon_{K_s}(z)$ from the same BC model.
Our stellar mass estimates are consistent with that listed in \citet{lin03b}, 
and are similar to that of \citet{andreon10} and \citet{gonzalez07}, when the intracluster light (ICL) contribution is excluded from the latter study.
For a few clusters at $z<0.1$ we have also used WISE data to estimate $M_{\rm star}$, and found agreement with 2MASS-based values to within 10\%.

\section{Correlations between Stellar Mass, ICM Mass, and Total Mass}
\label{sec:mass}

\begin{figure}
\epsscale{1.}
\plotone{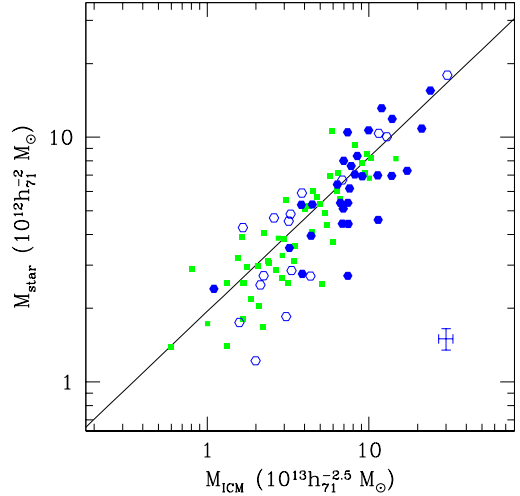}
\vspace{-4mm}
\caption{ 
Correlation between the total stellar mass and ICM mass, both measured within $r_{500}$, for the M08 clusters (filled circles), V09 clusters (open points), and $z<0.1$ clusters (green/squares).  
To avoid overcrowding the plot, errorbars for a typical cluster are shown in the lower right corner.  
We include a 10\% uncertainty in stellar mass for the assumption that a single value of mass-to-light ratio can describe the bulk of galaxies in a cluster.  
The choice of the IMF is the largest systematic uncertainty, at the 30\% level.
The line is the best-fit to clusters at $z\sim 0$ (Eqn.~\ref{eq:mgms}). 
}
\label{fig:allcls}
\end{figure}

In Fig.~\ref{fig:allcls} we show the $M_{\rm star}$--$M_{\rm ICM}$ correlation, measured within $r_{500}$, for the M08 clusters (filled circles), V09 clusters (open points), and $z<0.1$ clusters (green squares).  The M08 and V09 samples, although constructed using different selection criteria, appear to follow the same trend. 
Together the sample spans a factor of 30 in $M_{{\rm ICM}}$, allowing us to determine the slope of scaling relations well.

\begin{figure}
\epsscale{1}
\plotone{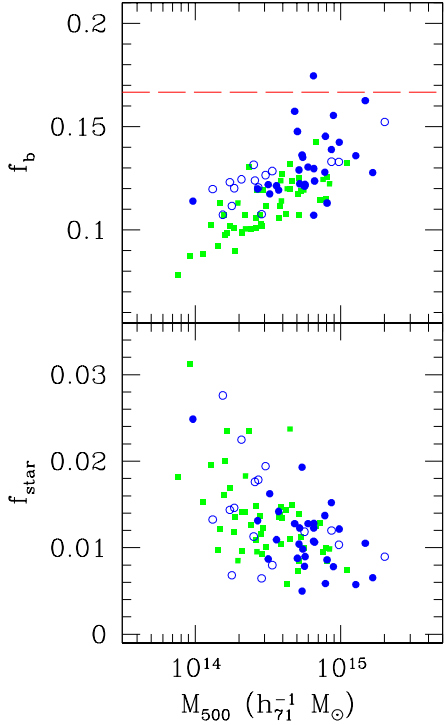}
\vspace{-7mm}
\caption{ 
The upper panel shows the baryon mass fraction, while the lower panel shows the stellar mass fraction.  The solid (open) circles are M08 (V09) clusters, and the solid squares are $z<0.1$ clusters.  The horizontal dashed line is the cosmic baryon fraction from {\it WMAP} \citep{komatsu09}.  
No attempt has been made to include the contribution from the ICL.
}
\label{fig:fb}
\end{figure}

The correlations between the baryon components and the cluster mass are shown in Fig.~\ref{fig:fb}.  The lower sets of points are the stellar mass fraction ($f_{\rm star} = M_{\rm star}/M_{500}$), and the upper sets are the baryon fraction, $f_{\rm b}=(M_{\rm ICM}+M_{\rm star})/M_{500}$, which is dominated by the ICM.

To investigate any redshift evolution among these correlations, we fit the data with the following form by $\chi^2$ minimization, taking into account the uncertainties in both mass measurements:
\begin{eqnarray}
\label{eq:3eq}
M_{\rm star}  & \propto & M_{\rm ICM}^{s_1} (1+z)^{\gamma_1} \nonumber\\ 
M_{\rm star}  & \propto & M_{500}^{s_2} (1+z)^{\gamma_2} \nonumber \\
M_{\rm ICM}  & \propto & M_{500}^{s_3} (1+z)^{\gamma_3}.
\end{eqnarray}
We find that $\gamma_1=-0.31 \pm 0.22$, $\gamma_2=-0.06\pm 0.22$, and $\gamma_3=0.41 \pm 0.14$; $s_1=0.62\pm 0.04$, $s_2=0.71 \pm 0.04$, and $s_3=1.13\pm 0.03$. 
That is, there is no evolution for the $M_{\rm star}$--$M_{500}$ relation, weak suggestion of evolution in $M_{{\rm star}}$--$M_{{\rm ICM}}$, and strong evidence of evolution in $M_{\rm ICM}$--$M_{500}$,
which is manifested as an offset between the loci of $f_{\rm b}$ for $z<0.1$ (solid squares) and $z>0.1$ (circles) clusters in Fig.~\ref{fig:fb}.  This is mainly due to the self-similar evolution (SSE) of the ICM, as noted first by V09 (see the discussion associated with Fig.~10 therein).  Let us denote $M_{\rm ICM} \propto M_{500}^{1+\kappa}$, where $\kappa \equiv s_3-1= 0.13$ based on our data. Introducing the nonlinear mass scale, $M_{\rm NL}$, 
satisfying $\sigma(M_{\rm NL}) D(z)=\delta_c\approx 1.69$, where $\sigma(M)$ is the {\it rms} fluctuation of linear power spectrum and $D$ is the growth factor \citep[][]{peebles80},
we can expect the ICM mass fraction to be the same for systems of the same $M_{500}/M_{\rm NL}$ at different redshifts under the SSE.  
It then follows that $M_{\rm ICM} \propto M_{500}^{s_3}/M_{\rm NL}^{\kappa}$.  
We note that $M_{\rm NL}(z)^{-\kappa}$ is 
consistent with
$(1+z)^{\gamma_3}$ up to $z\sim 0.4$ or so.
Indeed, with total mass scaled by $M_{\rm NL}$, the scatter reduces from 12\% to 8\%.
We have confirmed the consistency between the observed $M_{\rm ICM}$--$M_{500}$ evolution and the SSE expectation by repeating the above analysis using $M_{500}$ based on the $T_X$--$M_{500}$ relation of V09.

Using all of the clusters we find that
\begin{eqnarray}
\label{eq:mgms}
\frac{M_{{\rm star}}}{10^{12} h_{71}^{-2} M_\odot} & = & \left( 1.9 \pm 0.1 \right) \left( \frac{\Upsilon_{\rm 0,W1}}{1.0 \Upsilon_{\odot,{\rm W1}}} \right)  \\
 & & \mbox{}  \times \left( \frac{M_{{\rm ICM}}}{10^{13} h_{71}^{-2.5} M_\odot} \right)^{0.62 \pm 0.04} \left( 1+z \right)^{-0.31 \pm 0.22}  \nonumber
\end{eqnarray}
with a scatter of 31\%; Without strong evidence of evolution, for simplicity we take $\gamma_2 = 0$ in Eqn.~\ref{eq:3eq} and find that
\begin{equation}
\frac{M_{{\rm star}}}{10^{12} h_{71}^{-2} M_\odot} =  \left( 1.8 \pm 0.1 \right) \frac{\Upsilon_{\rm 0,W1}}{1.0 \Upsilon_{\odot,{\rm W1}}}  \left( \frac{M_{{\rm 500}}}{10^{14} h_{71}^{-1} M_\odot} \right)^{0.71 \pm 0.04} 
\end{equation}
with a scatter of 31\%.

The stellar mass-to-light ratio $\Upsilon$ is the most important systematic uncertainty in our results.
We adopt the Kroupa IMF, as it gives $\Upsilon$ at $z\sim 0$ that is consistent with the SAURON measurements of nearby elliptical galaxies \citep{cappellari06}.
Using the Salpeter (Chabrier) IMF, our $M_{\rm star}$ would be 34\% higher (24\% lower). 
Using a single value of $\Upsilon$ for all galaxies in a cluster is admittedly too simplistic \citep{leauthaud11}, although we note the spread of $\Upsilon$ in W1 for galaxies of all types is 20\% smaller than in optical bands. 
Furthermore, as galaxies in the $M^*\pm 1$ magnitude range contribute 65\% of the total light, our approach is reasonable, as long as $\Upsilon$ is only a weak function of galaxy mass (and/or morphology).

Another systematic uncertainty stems from the relative mass calibration of the $Y_X$--$M_{500}$ relation at different redshifts, which is about 5\% between $z=0$ and $z=0.5$ (V09).  
By boosting the total and ICM mass by 5\% for all clusters at $z>0.45$, we find that the exponents $\gamma_1$ and $\gamma_2$ become more negative (both change by $\sim 0.08$), giving a weak hint of changing in the stellar mass content.

If only using clusters at $z<0.4$, we would have found that $\gamma_2=-0.81 \pm 0.47$ (c.f.~$-0.06\pm 0.22$ when using the whole sample).  That is, the 17 highest-$z$ clusters have a substantial leverage on the determination of $\gamma_2$.  
It is also these clusters that exhibit ``brightening'' of galaxies with respect to the BC model.
Should we have adopted the measured $m^*$ based on these clusters (instead of using the BC model), 
which is equivalent to reducing $L_{\rm tot}$ and adopting a higher galactic stellar mass limit, we would have obtained 
$\gamma_2=-0.36 \pm 0.24$.
We therefore acknowledge the possibility that our results may be driven by the behavior of these most massive clusters at $z\sim 0.5$.
Ideally we would use a large, volume-limited sample that may represent the average cluster properties better, which will be carried out in a future publication with the all-sky WISE data.

Possible systematic effects aside,
the apparent lack of redshift evolution in $M_{\rm star}$--$M_{500}$ relation is consistent with the findings of G09 and L06.
In analyses that utilize $f_{\rm b}$ to infer cosmological parameters \citep[e.g.,][]{allen04}, the locally derived $M_{\rm star}$--$M_{500}$ relation is usually assumed to hold at higher redshifts.  Our result directly validates such an assmption.

Our results suggest that, within $r_{500}$, the gas and galaxy contents evolve in different ways; while gas mass grows according to the SSE fashion, the much larger scatter in $M_{\rm star}$--$M_{500}$ and $M_{\rm star}$--$M_{\rm ICM}$, as well as the much-less-than-unity slopes in these scaling relations, suggest a more stochastic growth history, which likely involves tidal interactions to strip off material from galaxies (e.g., L04, \citealt{conroy07}).  
In light of this, it would be critical to constrain the evolution of the stellar mass contained in the ICL. 
The upcoming Subaru {\it HyperSuprime Cam} survey \citep{takada10} will likely provide necessary data in this regard.
It is equally important to 
examine the baryon content evolution beyond $r_{500}$, both observationally and theoretically.

\acknowledgments

We are very grateful to 
G.~Bruzual and S.~Charlot for providing an updated version of their model,  to M.~Tanaka and C.~Mancone for help with the BC model predictions, and to an anonymous referee for very helpful comments.
YTL thanks E.~Komatsu, J.~Gunn, C.~Conroy, M.~Fukugita, M.~Takada, and D.~Spergel for helpful discussions, and I.H.~for constant encouragement.
YTL acknowledges supports from the World Premier International Research Center Initiative, MEXT, Japan.
This research was supported in part by the NSF under Grant No.~NSF PHY05-51164.
This publication makes use of data products from WISE, a joint project of UCLA and JPL/Caltech, funded by NASA,
and 2MASS, a joint project of UMass and IPAC/Caltech, funded by NASA and NSF.
Funding for SDSS-III has been provided by the Sloan Foundation, the Participating Institutions, NSF and DOE. 
This work makes use of data from CFHTLS and UKIDSS.

\end{document}